\documentstyle[prl,aps,emlines2]{revtex}

\begin{document}
\draft
\twocolumn[  
\title{  Reconstructing the density operator via simple projectors}
\author{ Ole Steuernagel }
\address{
  Arbeitsgruppe ``Nichtklassische Strahlung''
  der Max-Planck-Gesellschaft
  an der Humboldt--Universit\"at zu Berlin,\\
  Rudower Chaussee 5, 12484 Berlin, Germany, email: ole@photon.fta-berlin.de
}
\author{John A. Vaccaro}
\address{
  Physics Department, The Open University, Walton Hall, Milton Keynes MK7 6AA,
UK }
\date{Oct.,11,95}
\maketitle
\widetext   
\begin{abstract}
We describe the representation of arbitrary density operators
in terms of expectation values of simple projection operators.
Two representations are presented which yield
{\em non--recursive} schemes for experimentally
determining the density operator of any quantum system.  We
suggest a possible experimental implementation in quantum optics.

\end{abstract}
\pacs{03.65 Bz, 42.50.-p}
]   
\narrowtext 
\section{Introduction}
In the realm of quantum theory a state of a physical system is most generally
expressed by its density operator $ \hat \varrho $. Knowledge of this operator
gives complete information of the quantum state.
Schemes have been proposed in a number of fields in quantum physics to
determine $ \hat \varrho$ from measurements either explicitely
\cite{Fano57,Gale68,Park71,Royer85,Royer89,John&Ole,Smithey93}
or indirectly via quasiprobability distributions
\cite{Vogel89,D'Ariano94.a,Bohn91,Kuehn94}
for mixed states and also for pure states only
\cite{Freyberger94,Bardroff95}.

In this Letter we describe a general method of representing any
density operator $\hat \varrho$ in terms of expectation values
of simple projection operators.
Since the expectation values of projectors can, in principle,
be determined
experimentally this approach leads to
schemes for experimentally determining the
density operator.

Our approach differs from previously
proposed schemes in quantum optics for determining the density operator in its
use of
simple projectors
which project onto a single or a linear superposition of two basis
  states \cite{Gale68,Royer89}.
We place an emphasis on a 'minimalistic' representation
which comprises the least number of projection operators and
thus leads to the most efficient scheme, it is a generalization of the previous
considerations in \cite{Gale68,Royer89}.

The plan of the paper is as follows.
First we introduce the general idea of our approach, then,
in section II, we cast it into two specific representations and describe
their relative virtues.  In section III we describe a quantum optical
implementation and we end with a discussion in section IV.
%
%
\\

Let us assume that the Hilbert space representing the
states of the physical system is of countable
dimension $N$ and let $|m\rangle$ for
$m=1,...,N$  be {\em any conveniently chosen}
orthonormal basis of the space.
In cases where the space is infinite in dimension,
all expressions containing $N$ here and in the
following are infinite also.  Our
primary aim is to represent the $N^2-1$ independent
density matrix elements
$\varrho_{nm} \equiv \langle n | \hat \varrho | m \rangle $
in terms of the expectation values of simple projection operators.
Clearly the matrix elements cannot be
expressed solely in terms of
the {$N-1$} independent
expectation values $\varrho_{mm}=\langle|m\rangle\langle m|\rangle$
of the set of the $N$ base state projectors $|m \rangle\langle m|$
because the vital phase
information of the coherences, i.e. the complex nature of
the off-diagonal elements $\varrho_{nm}$ for $n\ne m$,
can not be derived from the diagonal elements alone.

The simplest possible generalization of the base state
projectors is given by the set of
projection operators which project onto linear combinations
of {\em two} base states,
e.g. onto
{$c_1 |n\rangle+c_2 |m\rangle$.}
The expectation value of such
projectors represent the phase information of the
coherences in its most elementary form. We show that
{\em one can express $ \hat \varrho $  in
terms of expectation values of such projectors
and how to implement it quantum optically}.
\section{Two representations}

For simplicity let us consider the $\{ | n \rangle , | m \rangle  \}$--subspace
which is spanned by any two basis vectors $|n\rangle$ and $|m\rangle$ for $n\ne
m$
and define the state
\begin{equation}
 | a \rangle  \equiv N_a ( | n \rangle  + a | m \rangle  ) \; ,
\label{musterzustand}
\end{equation}
where  $N_a = 1/\sqrt{1 + |a|^2}$ is a normalization
constant  and $a \equiv |a|e^{  {\rm i} \alpha}$ is a nonzero coefficient.
A corresponding
nomenclature is used for a second, different state of the
same subspace $ | b \rangle  = N_b (| n \rangle  + b | m \rangle  ) $, where
$ b = |b| e^{ {\rm i}\beta}\ne a $. We defer making any further restriction on
the values of $a$ and $b$,
to guarantee independence of the expectation values of the corresponding
projectors
\begin{eqnarray}
\nonumber
\hat A & \equiv & | a \rangle \langle a | \ , \\
\hat B & \equiv & | b \rangle \langle b | \ ,
\label{defnAB}
\end{eqnarray}
until  later.

Let us assume the measurements yielding
the expectation values of the projectors
$|n\rangle\langle n|$, $|m\rangle\langle m|$, $\hat A$ and $\hat B$
have been performed
{\cite{specialcase}
}. The first two expectation values are simply the diagonal elements
$\varrho_{nn}$, $\varrho_{mm}$.
We can combine these expectation values conveniently as
\begin{eqnarray}
 M_{| a\rangle}  & \equiv & {\mbox {Tr} } \{  \hat \varrho \hat A \} - N_a^2
(\varrho_{nn} + |a|^2 \varrho_{mm})
\nonumber \\
& = & N_a^2 ( a \varrho_{nm} + a^\ast \varrho_{mn} ) ,
\label{MMessung}
\end{eqnarray}
where Tr is the trace operation and $ M_{| a\rangle} $ stands for
the result associated with a
measurement of the projector $\hat A$.
A corresponding expression is obtained
for the result $ M_{| b \rangle} $ associated with the projector
{$\hat B $
}.
Let us write $\varrho_{nm}$
in terms of its real and imaginary parts $\varrho_{nm}\equiv R +  {\rm i} J$
and let us define
\begin{eqnarray}
  m_{|a\rangle}  & \equiv & \frac{M_{| a \rangle}}{2 |a| N_{a}^2}   = R \cos
\alpha - J \sin \alpha \; ,
\nonumber \\
\mbox{and  } \quad  m_{|b\rangle} & \equiv &
  \frac{M_{| b \rangle}}{2 |b| N_{b}^2}  = R \cos \beta - J \sin \beta \; .
\label{measure}
\end{eqnarray}
Solving these equations for $R$ and $J$ yields
\begin{eqnarray}
\left(
\begin{array}{c} R \\ J \end{array}
\right) & = &
\frac{ 1 }{ \sin (\beta -\alpha) }
\left(
\begin{array}{cc} \sin \beta & - \sin \alpha \\
\cos \beta & - \cos \alpha \end{array}
\right)
\left(
\begin{array}{c} m_{|a\rangle}  \\ m_{|b\rangle} \end{array}
\right)
\nonumber \\
& \equiv & \mbox{\boldmath$ T $} \left(
\begin{array}{c} m_{|a\rangle}  \\ m_{|b\rangle} \end{array}
\right)  \; .
\label{loesung2}
\end{eqnarray}
Clearly this requires $\beta - \alpha
\neq k \pi $, where $k$ is any integer.
This gives the only restriction on the values of $a$ and $b$ aside from the
trivial requirement that $a \neq 0 \neq b$.
Applying the outlined procedure to the
$\{ | n \rangle , | m \rangle  \}$--subspaces
for $1\le n<m\le N$
allows us to represent $\hat \varrho$
in terms of expectation values of $N^2-1$ different projectors,
{due to the condition Tr $\hat \varrho = 1$}. Note that this scheme
is intrinsically {\em non-recursive}.

We call this the 'minimal' representation as it requires this least
possible number of projection operators to represent a general
density operator and also because it puts almost no restrictions
on the states forming the projectors, namely on the coefficients
$a$ and $b$ of Eq.\ (\ref{musterzustand}).
\\

Though mathematically satisfactory the  minimal
representation would be sensitive to experimental
errors in a physical implementation.
This sensitivity however is minimized using {\em sensitivity
optimized states}, i.e. choosing $|a|=|b|=1$
and $b=\pm $ i$\; a$ \cite{sensitivity}.
This sensitivity can be further reduced employing three
or more (redundant) states. Let us for example look at the case of
one more
projector state $|c\rangle \equiv N_c (|n\rangle+c|m\rangle)$
where $c=|c|e^{{\rm i}\gamma }$ in each $\{n,m\}$--subspace.
We find that
\begin{equation}
  m_{|c\rangle} = \frac{ m_{|a\rangle}  \sin(\beta-\gamma ) -
    m_{|b\rangle} \sin(\alpha-\gamma )}{\sin(\beta - \alpha)} \; ,
\label{xrel}
\end{equation}
where $m_{|c\rangle}$ is given by Eqs.\ (\ref{measure},\ref{MMessung})
with $|b\rangle$ replaced with $|c\rangle$.
Provided the differences {$\alpha-\gamma $, $\beta-\gamma $
and $\alpha-\beta$} between the phase angles of the states
$|a\rangle$, $|b\rangle$ and $|c\rangle$ are
{not multiples}
of $\pi$ the overparameterization introduced by the
extra state can be used to reduce the effect of experimental errors.
For example, one could estimate true values of
$m_{|a\rangle}$, $m_{|b\rangle}$ and $m_{|c\rangle}$ as the point
$(x,y,z)$ on the surface
$z(x,y)=[x\sin(\beta-\gamma ) - y\sin(\alpha-\gamma )]/\sin(\beta-\alpha)$
which is closest
to the point $(\bar x,\bar y,\bar z)$ where
$\bar x$, $\bar y$ and $\bar z$ are the experimentally
measured values of $m_{|a\rangle}$, $m_{|b\rangle}$ and
\nolinebreak $m_{|c\rangle}$.
\\

One may still go one step further and consider the particular quadruplet
of states
\begin{eqnarray}
 | a^{nm}_{\pm} \rangle  &\equiv& \frac{1}{\sqrt{2}} ( | n \rangle  \pm | m
\rangle  ) \;  , \nonumber \\
 | b^{nm}_{\pm} \rangle  &\equiv& \frac{1}{\sqrt{2}} ( | n \rangle  \pm  {\rm
i} | m \rangle  ) \;  ,
\label{quadrupel}
\end{eqnarray}
for {$ n,m=1,2,\dots,N$.}
We mention in passing that all such states are normalized except for
$n=m$ for which $|a^{nn}_+\rangle \equiv \sqrt{2}|n\rangle$ and
$|a^{nn}_-\rangle , |b^{nn}_\pm \rangle \equiv 0$.
The set $\{ | a^{nm}_{\pm} \rangle , | b^{nm}_{\pm} \rangle:$
$ m,n=1,...,N \}$
is an overcomplete basis of the Hilbert space.
Let the projection operators
\cite{projector}
which project onto these states be
$ \hat A^{nm}_{\pm}  $ $\equiv $
$ | a^{nm}_{\pm} \rangle   \langle { a^{nm}_{\pm}} |$,
$ \hat B^{nm}_\pm $$\equiv $$ | b^{nm}_{\pm} \rangle   \langle { b^{nm}_{\pm}}
| $,
{defined in analogy to Eq.\ (\ref{defnAB}).}
The expectation values of
the $2N^2-N$ different projectors \cite{2n2n} for $n\le m$
suffice to represent an arbitrary matrix element of $\hat \varrho$ as
\begin{equation}
\label{fano_rep}
\varrho_{mn}  \; = \;  {\mbox {Tr} } \{  \hat \varrho
\; \frac{1}{2} [ \hat A^{nm}_+  -  \hat A^{nm}_-  +  {\rm i}
( \hat B^{nm}_+ -  \hat B^{nm}_- ) ] \} \; ,
\end{equation}
a form that has already been derived in \cite{Gale68,Royer89}.
Now the projectors can be combined to form operators
$\hat R^{nm}$, $\hat J^{nm}$ defined as
\begin{eqnarray}
&& \hat R^{nm} \equiv (  \hat A^{nm}_+  -  \hat A^{nm}_-  )/\sqrt{2} \;
= ( |n \rangle \langle m | + |m \rangle \langle n | )/\sqrt{2} \;  ,
\nonumber \\
&& \hat J^{nm} \equiv (  \hat B^{nm}_+ -  \hat B^{nm}_-  )/\sqrt{2} \;
= {\rm i} ( |n \rangle \langle m | - |m \rangle \langle n | )/\sqrt{2} \; ,
\label{randj}
\end{eqnarray}
fulfilling the orthogonality relations
\begin{eqnarray}
\nonumber
&&  { \mbox {Tr} } \{  \hat R^{nm} \hat R^{pq} \}  \; = \;  ( \delta_{n,p}
\delta_{m,q} +
\delta_{n,q} \delta_{m,p} ) \;  , \nonumber \\
&&  {\mbox {Tr} } \{  \hat J^{nm} \hat J^{pq} \}  \; = \;  ( \delta_{n,p}
\delta_{m,q} -
\delta_{n,q} \delta_{m,p} ) \;  , \nonumber \\
& \mbox{ and } &  {\mbox {Tr} } \{  \hat R^{nm} \hat J^{pq} \}  \; = \;  0 \;
,
\end{eqnarray}
for $n,m,p,q = 1,...,N$,
where $\delta_{n,m}$ is the Kronecker delta.
The set $\{\hat R^{mn},\  \hat J^{mn}\ :\  n \leq m \}$
constitutes a complete basis set of $N^2$ operators.
This operator basis gives an unique expansion
of any operator $\hat Q$ as
\begin{eqnarray}
\nonumber
\hat Q  \; = \;  \sqrt{2}\big(
\sum_{m=2}^{N}\sum_{n=1}^{m-1} r_{nm} \hat R^{mn}
+ j_{nm} \hat J^{mn}  \big) \\
+  \frac{1}{\sqrt{ 2 } }\sum_{m=1}^{N} r_{mm} \hat R^{mm} \; ,
\label{genexp}
\end{eqnarray}
with $ r_{nm} = {\mbox {Tr} } \{ \hat Q \hat R^{mn} \}/\sqrt{2}
= (Q_{mn} + Q_{nm})/2$ and
$ j_{nm}
= {\mbox {Tr} } \{ \hat Q \hat J^{mn} \}/\sqrt{2} =
(Q_{mn} - Q_{nm}) $i$/2$.
If $\hat Q$ is a hermitian operator $r_{nm}$ and $j_{nm}$
are the real and imaginary parts of the matrix elements
$Q_{nm}\equiv \langle n | \hat Q | m \rangle $.
\\

Fano introduced the idea of expanding the density matrix in terms of an
orthogonal
operator basis \cite{Fano57}, hence we call this an 'operator basis'
representation. We
introduced this representation for its mathematical properties rather than
its physical contents.
Let us note that the sensitivity optimized states mentioned before
Eq.(\ref{xrel})
can analogously be cast into this kind of orthogonal operator basis, in this
sense the operator basis representation is contained in the minimal one.
\section{Quantum Optical realization}
Next we describe a possible experimental scheme for the reconstruction
of a density operator describing the state of a single optical field mode
{\cite{underpins}.}
It is a straightforward matter to generalize this to several optical modes.
{We use} the {\em Fock state basis} in which the numbers of
photons in the mode under consideration label the states
$\{ | m \rangle : m = 0,1,2,...\}$.
Our task is to show that the expectation values of the
{corresponding} projection
operators $\hat A^{nm}_\pm$, $\hat A$, etc. can be obtained
experimentally.
We note from the outset that
the experimentally difficult part
of the scheme at present is the preparation of
coherent superpositions of two Fock states.
However, in the light of recent theoretical \cite{Kilin95,coherentfockstates}
and experimental results \cite{An95}, it is clear that the problem of the
preparation of the probe field can and will be solved.

Thus, since this is
not a {\em fundamental} difficulty we assume in the following that
such superposition states are available.

The expectation value of the projection operators
in the representations
can be determined using the experimental setup depicted
in Fig.\ 1 as follows.
\begin{figure}
\unitlength=0.55mm
\special{em:linewidth 0.8pt}
-------------------------------------------------------------------\\
If you are interested in the figure, please contact me via email or fax
and send your fax number, then I'll send the picture by fax.
\\
Ole Steuernagel \\
email : ole@photon.fta-berlin.de \\
fax: ++49 (germany) -30- 6392 3990 \\
--------------------------------------------------------------------\\
\caption{
The setup of our proposed quantum optical scheme. Light from a common field
source
is fed into a device generating the probe field $| a \rangle \langle a |$ and a
device
that generates the signal field $\hat \varrho$.  The probe and signal fields,
which are labelled 1 and 2, respectively, are then entangled
at the last beam splitter and analyzed by the photodetectors $I$ and $II$.
The use of a common
source ensures that the probe and signal fields oscillate at the same
frequency.
}
\label{fig1}
\end{figure}
A probe field is prepared in a particular state $|\psi\rangle $
and fed into port 1 of the beam splitter, the signal
field prepared in the (unknown) state $\hat \varrho$ is fed
into port 2.  The joint photon number probability distribution
of the output ports of the beam splitter
is obtained from the photoelectron
statistics produced in the photodetectors $I$ and $II$
for many repetitions of the experiment, let us note that multiphoton
coincidence counts together with quantum efficiencies above 70\% have been
demonstrated experimentally \cite{Petroff87,Kwiat93pra48}.
If one chooses a method that detects single photons with more than 50\%
quantum efficiency the photon number probability distribution can be recovered
from the measurements using the inverse Bernoulli transformation discussed
by Lee \cite{lee}.

Furthermore a
new method developed by Munroe {\em et al.} \cite{Munroe95} allows to
measure the photon-number statistics from the phase-averaged quadrature-field
distribution with single photon and ultrahigh time resolution of the order of
300 fs. Employing the corresponding reconstruction schemes
\cite{Leonhardt95reco}
this method yields
almost perfect photon number statistics.

Hence we may restrict our considerations to the
'true' joint photon probability distribution $P_{|\psi\rangle}(p,q)$ for $p$
and $q$
photons measured by (ideal) photodetectors $I$ and $II$,
respectively, which is given by
\begin{eqnarray}
\nonumber
& &P_{|\psi \rangle}(p,q) =
 \sum\limits^{p+q}_{n'=0} \sum\limits^{p+q}_{m'=0}
   \langle n'|\hat \varrho| m'\rangle \langle p+q-n'|\psi\rangle\langle\psi|
p+q-m'\rangle\\
& &\ \ \ \times A_p(n',p+q-n') A^\ast_p(m',p+q-m')\ .
\label{eqn4}
\end{eqnarray}
Here $A_p(\nu ,\mu )$ represents the probability
amplitude of finding mode $I$ in the Fock state $|p\rangle_I$
if modes $1$ and $2$ are in the product
Fock state $|\nu \rangle_1 |\mu \rangle_2 $ and is given by
\begin{eqnarray}
\nonumber
& &A_p(\nu ,\mu)= (-1)^\nu  \sqrt{\frac{p!(\nu +\mu-p)!}
	       {\nu ! \mu! }}
	   \; {\rm e}^{{\rm i}\varphi_\tau(p-\mu)}
	   \; {\rm e}^{{\rm i}\varphi_\rho(p-\nu )}\\
& &\ \ \times \sum\limits^\nu _{k=0}\sum\limits^\mu_{l=0}
	   (-1)^{k}{\nu \choose k}{\mu\choose l}
	   \sqrt{\tau^{\mu+k-l}\rho^{\nu -k+l}} \; \delta_{k+l,p} \; ,
\label{eqn2}
\end{eqnarray}
where $\tau$, $\rho$ are the transmittance and reflectance and $\varphi_\tau$,
$\varphi_\rho$ are the corresponding phase factors generated by the beam
splitter
as defined by Campos {\em et al.} in Ref.\ \cite{Campos89}.
Inserting for $|\psi \rangle $ the special probe field states
$ | a^{nm} \rangle $ with $n>m$,
see Eq.\ (\ref{musterzustand}),
and relabeling $ p + q = N + n = M + m $ changes Eq.\ (\ref{eqn4}) into
\begin{eqnarray}
  P_{ | a^{nm} \rangle  }(p,N +\, n-p) & = &
C \langle a' | \hat \varrho | a' \rangle \; ,
\label{photoschema}
\end{eqnarray}
where
\begin{equation}
\begin{array}{rll}
 | a' \rangle  = &  C^{-1/2} & [  A_p^\ast(N,n)
 N_a |N \rangle +  A_p^\ast(N,n)  N_a a^\ast |M \rangle ] \qquad
\nonumber \\
\mbox{and } \qquad
& C  = & | A_p(N,n) N_a |^2 + | A_p(N,n)  N_a a |^2 \; .
\label{photoschemaext}
\end{array}
\end{equation}
Again we assume that the diagonal elements are known, for example by the
comparatively simple
measurement of the photo count distribution of the field alone.
The same is assumed to be true for $a$, which is known from the state
preparation process, we can thus,
equivalently to Eq.\ (\ref{MMessung}),
use $ P_{ | a^{nm} \rangle  }(p,N +\, n-p)$ to determine a quantity
\begin{eqnarray}
M_{  | a^{nm} \rangle  }( N ,p) \equiv &&
2  \mbox{Re} \{ a \; \varrho_{MN} \; A_p(M,m) A^\ast_p(N,n) \} \; ,
\label{nix}
\end{eqnarray}
where 'Re' signifies the real part. Using a second linearly independent probe
state $ | b^{nm} \rangle  $, by a procedure analogous to Eqns.\ (\ref{measure})
and (\ref{loesung2}) we obtain $ \varrho_{MN} $.
Thus {\em we have translated the minimal representation into an experimental
scheme in quantum optics for determining the quantum state of light;}
the translation of the other representations along similar lines is
straightforward.
\\

It is interesting to note that the value of $p$
in Eq.\ (\ref{photoschema}) can be chosen arbitrarily from the interval
$(0 \leq p \leq N+n)$.  This gives $N+n+1$ different ways of
determining the value of the quantity
$M_{  | a^{nm} \rangle  }( N ,p) $ in Eq.\ (\ref{nix}).
Also, since we require $n-m=M-N$ in Eq.\ (\ref{photoschema})
the set of matrix elements $\varrho_{(k+N-M) \; k}$ for $k=0,1,2,...$
can be determined from just two probability distributions
$ P_{ | a^{nm} \rangle  }$ and $ P_{ | b^{nm} \rangle  }$
for fixed values of $n$ and $m$.
And finally, since it is the difference $n-m$ only that decides
which set of matrix elements are determined
this implementation is also redundant in the sense that
the probe states $| a^{s t} \rangle$ with $s=t+n-m$
are equivalent for $t=0,1,2,...$.

This scheme will give as many matrix elements
of the density operator as desired and is limited only by experimental
error and the ability to prepare the probe field in suitable two-Fock
state superpositions.
\section{Discussion}
We examined the requirements for representing any density operator
in terms of expectation values of simple projection operators.
We gave two different representations:
the minimal representation which
requires the least number of projectors and the
operator basis representation
which gives the expansion of any operator
in terms of an operator basis.
Our results are applicable to any physical system whose
state space is of countable dimension $N$ which
need not be finite.

We showed how the expectation values could be determined
experimentally  for the case of a single mode of an optical field.
An important point about our method is that it is {\em not recursive},
in contrast to some other methods
for determining the density operator of the optical field
\cite{John&Ole,Freyberger94,Bardroff95}
for which the calculation of all but a select
few matrix elements involves the previously calculated
values of other matrix elements and results in the accumulation of
experimental errors.
%
%
\\\\
This work was supported by Max--Planck--Gesellschaft.
Ole Steuernagel thanks Harry Paul, Yacob Ben--Aryeh, Robert Lynch, T\'amas
Kiss,
Ulf Leonhardt, and Mladen Pavi\v ci\'c for clarifying discussions.
\\ \\
%



\begin{references}
\bibitem{specialcase} Clearly $|n\rangle \langle n|$ and $ |m\rangle \langle
m|$
are only special cases of the above mentioned projectors in Eq.\
(\ref{defnAB}).
\bibitem{sensitivity} The first constraint results from the requirement
that $ m_{| a \rangle }$ in Eq.\ (\ref{measure}) shall be least sensitive
to variations of $a$, i.e. $\frac{\partial }{\partial a} (a N^2_a)^{-1} =0$,
and the same for $b$. The second stems from the requirement that the
overall multiplication factor of the vector $( m_{|a \rangle }
, m_{| b \rangle} )$,
 i.e. the
determinant of $\mbox{\boldmath$ T $}$ in Eq.\ (\ref{loesung2}), shall be least
sensitive to variations of the relative phase $\beta -\alpha $.
This implies $\beta - \alpha $ to be odd multiples of $ \pi / 2$
or $b = \pm $ i$\; a$.
\bibitem{projector} A projector is defined as $\hat P^2=\hat P$ and so
stricktly speaking
$\hat A^{nn}_+/2$ is a projector but $\hat A^{nn}_+$ is not.
However, for brevity, we also refer to the latter
as a projector in this paper.
\bibitem{2n2n}
These are $N + \sum_{n < m}^{N} 4 = N + 2 N (N-1)$
projectors to represent the diagonal and off--diagonal
terms of $\hat \rho$.
\bibitem{underpins}
Gale's proposed experimental scheme \cite{Gale68} for specifically
determining the density operator of a spin using combinations
of Stern--Gerlach spin filters is another possible experimental
realization of the representations presented above.
\end{references}
\end{document}